\documentstyle[twocolumn,epsfig]{jpsj}

\def\runtitle{}
\def\runauthor{Kouji {\sc Ueda}, Ryota {\sc Otani}, Yukinobu {\sc Nishio},
Andrej {\sc Gendiar}, and Tomotoshi {\sc Nishino}}

\title{Critical Point of a Symmetric Vertex Model}

\author
{Kouji {\sc Ueda}, Ryota {\sc Otani}, Yukinobu {\sc Nishio},
Andrej {\sc Gendiar}$^{2)}$, and Tomotoshi {\sc Nishino}}

\inst
{Department of Physics, Faculty of Science, Kobe University, Kobe 657-8501\\
$^{2)}$ Institute of Electrical Engineering, Slovak Academy of Sciences, D\'ubravsk\'a cesta 9, 
SK-841 04 Bratislava, Slovakia}

\begin{document}
\sloppy
\maketitle

The vertex models have been intensively studied and their thermodynamic 
properties are well known inside the solvable parameter area.~\cite{Lieb,Baxter} 
On the other hand, outside the solvable area the model is not
fully analyzed. Takasaki et al. numerically investigated a vertex model, which 
allows 7 vertex configurations and contains 2 parameters.~\cite{Takasaki} They observed 
a phase transition that belongs to the Ising universality class.
In this article we study a symmetric vertex model, that allows 10 vertex 
configurations, by use of the corner transfer matrix renormalization 
group (CTMRG),~\cite{CTMRG} 
a variant of the density matrix renormalization group.~\cite{DMRG,DMRG2,DMRG3} 
As we report in the following the model has a critical point. 

\begin{figure}
\epsfxsize=40mm 
\centerline{\epsffile{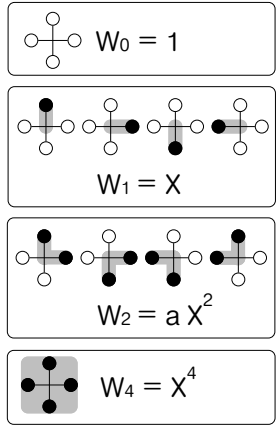}}
\caption{Allowed vertex configurations and their Boltzmann weights. 
The open and black circles, respectively, denote spins that take 0 and 1. 
The shaded area will be used when we plot the spin configuration 
snapshots in Fig.2 and 5.}
\label{fig:1}
\end{figure}

Let us introduce the vertex model that we study in the following. 
Consider a square lattice, where there is a 2-state spin variable 
 ($\sigma = 0$ or $1$) on each bond. Thus a lattice point is surrounded by
4 spins. We impose a local constraint for these 4 spins as shown in Fig.1.
We assign the following Boltzmann weights
\begin{equation}
W_0^{~} = 1         \, , \,\,\,\,
W_1^{~} = x         \, , \,\,\,\,
W_2^{~} = a x^2_{~} \, , \,\,\,\,
W_4^{~} =   x^4_{~}
\end{equation}
for these 10 local configurations, 
where the subscripts denote the sum of 4 spin variables. The parameter $x$ 
is positive. The expectation value of a spin
\begin{equation}
\rho( x, a ) \equiv \langle \sigma \rangle = ( P_1^{~} + 2 P_2^{~} + 4 P_4^{~} ) / 4 \, ,
\end{equation}
where $P_{\ell}^{~}$ is the probability to observe vertices with ${\ell}$ numbers of 1,
is the increasing function of $x$.

\begin{figure}
\epsfxsize=40mm 
\centerline{\epsffile{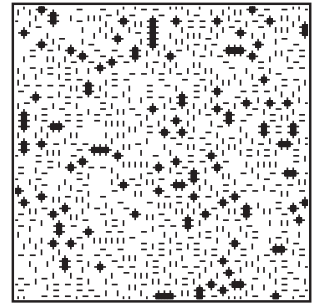}}
\caption{Spin configuration snapshot at $a = 0$ and $x = 1.23$. For each
vertex, the shaded area in Fig.1 is drawn. This configuration snapshot is 
obtained by recently developed extension of CTMRG.~\cite{Snap,Snap2}}
\label{fig:2}
\end{figure}

When the parameter $a$ is $0$, only 6 vertex configurations are allowed.
Under this strong constraint, rectangular areas inside which all spins
are $\sigma = 1$, appear separately in the `sea' of $\sigma = 0$, as 
shown in Fig.2. The expectation value $\rho( x, a = 0 )$ 
gradually increases with $x$ till $x = 1.23$. At this point
the system shows first order phase transition, and $\rho$
jumps to unity. (See Fig.3.)

\begin{figure}
\epsfxsize=65mm 
\centerline{\epsffile{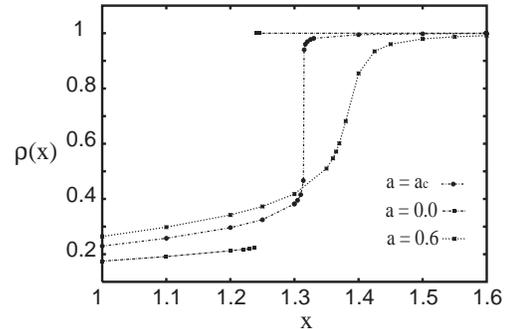}}
\caption{Expectation value $\rho( x, a )$ for $a = 0$, $0.3599$, 
and $0.6$.}
\label{fig:3}
\end{figure}

In contrast, when $a = 0.6$ there is no discontinuity nor singularity in 
$\rho( x, a = 0.6 )$. Thus there should be a critical point at certain 
values of $a$ and $x$. We trace the line of the first order transition in the
parameter space as shown in Fig.4. The critical value of $a$ is roughly
determined as 0.36. We further perform scaling analysis for $\rho( x, a )$,
and find one of the parameter line
\begin{equation}
a = - 0.2736 ( x - x_{\rm C}^{~} ) + a_{\rm C}^{~}
\end{equation}
shown in the inset, that passes the critical point at 
$( x_{\rm C}^{~}, a_{\rm C}^{~} ) = ( 1.31438, 0.3599 )$. 
Figure 5 shows the configuration snapshot at the criticality. 

\begin{figure}
\epsfxsize=65mm 
\centerline{\epsffile{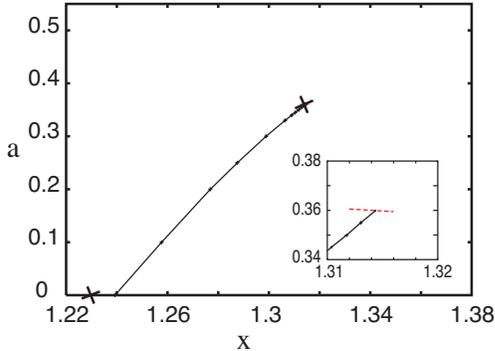}}
\caption{Phase Diagram. The line of the first order phase transition ends
at the critical point $( x_{\rm C}^{~}, a_{\rm C}^{~} ) = ( 1.31438, 0.3599 )$.
Cross marks represent data points used for the snapshot observation in Fig.2 
and Fig.5. Inset: the parameter line used for the scaling analysis in Fig.6.}
\label{fig:4}
\end{figure}

\begin{figure}
\epsfxsize=40mm 
\centerline{\epsffile{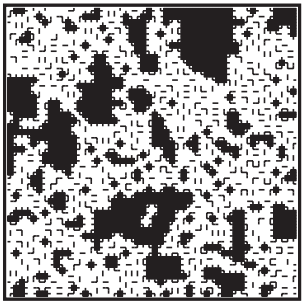}}
\caption{Snapshot at the criticality.}
\label{fig:5}
\end{figure}

In order to determine the spin expectation value 
$\rho( x_{\rm C}^{~}, a_{\rm C}^{~} )$ at the criticality, we
assume a scaling form 
\begin{equation}
\rho( x, a ) - \rho( x_{\rm C}^{~}, a_{\rm C}^{~} ) \propto
\left| x_{\rm C}^{~} - x \right|^\mu_{~} 
\end{equation}
on the parameter line that passes through the critical point,~\cite{tangent}
and plot $\rho( x, a )$ on the parameter line given by Eq.(3) with respect to a 
various power of $| x_{\rm C}^{~} - x |$. As a result we confirm that  $\rho( x, a )$ 
shows linear dependence with $| x_{\rm C}^{~} - x |^{1/15}_{~}$ as shown in Fig.6, 
where the two fitted lines are represented by the following equations
\begin{eqnarray}
 0.369 ( x - x_{\rm C}^{~} )^{1/15}_{~}  &+& 0.716 
 \,\,\,\,\,\,\,\,\,\, ( x > x_{\rm C}^{~} ) \nonumber\\
-0.422 ( x_{\rm C}^{~} - x )^{1/15}_{~}  &+& 0.716 
 \,\,\,\,\,\,\,\,\,\, ( x < x_{\rm C}^{~} ) \, .
\end{eqnarray}
In conclusion the exponent $\mu$ in Eq.(4) is $1/15$ and the value
$\rho( x_{\rm C}^{~}, a_{\rm C}^{~} )$ is $0.716$. Thus the 
critical point belongs to the same universality class as the 2D Ising model, 
whose magnetization at critical temperature obeys the scaling form
\begin{equation}
M( T_{\rm C}^{~}, h ) \propto | h |^{1/\delta}_{~}
\end{equation}
where $\delta = 15$.

\begin{figure}
\epsfxsize=65mm 
\centerline{\epsffile{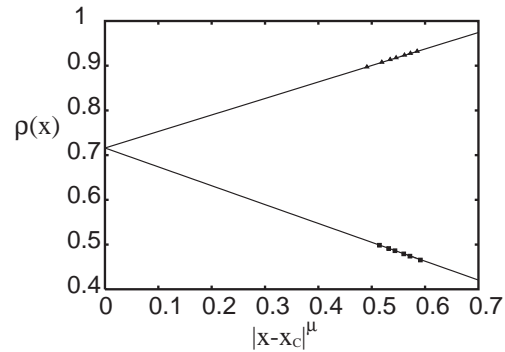}}
\caption{Spin expectation value $\rho$ on the parameter line given by Eq.(3). 
The exponent $\mu$ is fixed to $1/15$.}
\label{fig:6}
\end{figure}

To summarize, we investigated the phase transition of a vertex model 
with 10 local configurations, and found a critical point that belongs
to the Ising universality class. A possible extension of the current 
study is toward the direction of the lattice polymer.\cite{Polymer,Polymer2} For example,
introducing two new vertices to those vertices shown in Fig.1, 
one obtains a unified model with the 7-vertex case studied by Takasaki
 et al. that corresponds to a straight line polymer
in two dimension.~\cite{Takasaki} 

The authors thank to E.~Kaneshita for valuable discussions.
A.G. is supported by the VEGA grant No. 2/3118/23, and also by
Japan Society for the Promotion of Science (P01192).


\begin{thebibliography}{99}
\bibitem{Snap2} For the calculation to obtain Fig.2 and 5 we keep $m = 20$ block spin
states, the number which is sufficient for drawing snapshot of the size $50$ by $50$.
\end{thebibliography}
\end{document}